\documentclass{aa}  

\usepackage{graphicx}
\usepackage{txfonts}
\begin{document}

   \title{Stellar collisions in flattened and rotating Pop. III star clusters}
   
    \author{M.Z.C. Vergara$^{1,\star}$, D.R.G. Schleicher$^{1}$, T.C.N. Boekholt$^{2}$, B. Reinoso$^{3}$, M. Fellhauer$^{1}$, R.S. Klessen$^{3,4}$ $\&$ N. W. C. Leigh$^{1,5}$}
    \institute{$^{1}$ Departamento de Astronom\'ia, Facultad Ciencias F\'isicas y Matem\'aticas, Universidad de Concepcion, Av. Esteban Iturra s/n Barrio Universitario, Casilla 160-C, Concepcion, Chile
       \\$^{2}$ Rudolf Peierls Centre for Theoretical Physics, Clarendon Laboratory, Parks Road, Oxford, OX1 3PU, UK 
       \\$^{3}$ Universität Heidelberg, Zentrum für Astronomie, Institut für Theoretische Astrophysik, Albert-Ueberle-Str. 2, 69120 Heidelberg,
       Germany
       \\$^{4}$ Universität Heidelberg, Interdisziplinäres Zentrum für Wissenschaftliches Rechnen, Im Neuenheimer Feld 205, 69120 Heidelberg,
    Germany
       \\$^{5}$Department of Astrophysics, American Museum of Natural History, New York, NY 10024, USA
       \\$^{\star}$ marccortes@udec.cl}
   
   \date{Received February 22, 2021; accepted March 8, 2021}
   \titlerunning{Impact of flattening and rotation in Pop. III star cluster}
   \authorrunning{M. Z. C. Vergara et al.}

 
   \abstract
   {
   Fragmentation often occurs in disk-like structures, both in the early Universe and in the context of present-day star formation.  Supermassive black holes (SMBHs) are astrophysical objects whose origin is not well understood; they weigh millions of solar masses and reside in the centers of galaxies. An important formation scenario for SMBHs is based on collisions and mergers of stars in a massive cluster with a high stellar density, in which the most massive star moves to the center of the cluster due to dynamical friction. This increases the rate of collisions and mergers since massive stars have larger collisional cross sections. This can lead to a runaway growth of a very massive star which may collapse to become an intermediate-mass black hole. Here we investigate the dynamical evolution of Miyamoto-Nagai models that allow us to describe dense stellar clusters, including flattening and different degrees of rotation. We find that the collisions in these clusters depend mostly on the number of stars and the initial stellar radii for a given radial size of the cluster. By comparison, rotation seems to affect the collision rate by at most $20\%$. For flatness, we compared spherical models with systems that have a scale height of about $10\%$ of their radial extent, in this case finding a change in the collision rate of   less than $25\%$. Overall, we conclude that the parameters only have  a minor effect on the number of collisions. Our results also suggest that rotation helps to retain more stars in the system, reducing the number of escapers by a factor of $2-3$ depending on the model and the specific realization. After two million years, a typical lifetime of a very massive star, we find that about $630$ collisions occur in a typical models with $N=10^4$, $R=100$ $\rm~R_\odot$ and a half-mass radius of $0.1$ $\rm~pc$, leading to a mass of about $6.3\times10^3$ $\rm~M_\odot$ for the most massive object.
    We note that our simulations do not include mass loss during mergers or due to stellar winds. On the other hand, the growth of the most massive object may  subsequently continue, depending on the lifetime of the most massive object.}

   \keywords{cosmology:  theory — early Universe - dark ages, reionization, first stars - stars: Population III}

   \maketitle
%

\section{Introduction}

 Supermassive black holes
(SMBHs) are very compact astrophysical objects of  unknown origin weighing millions of solar masses ($ \sim 10 ^ 6-10 ^ 9~$M$_{\odot} $) and residing in the centers of galaxies \citep{V10}. The first image confirming the existence of these monsters was recently   observed with the Event Horizon Telescope \citep{EHT19}. Studies of stellar orbits have shown that there is a very compact object,
called Sagittarius A*, in the center of our own Galaxy, the Milky Way \citep{G08,G10}. The observed correlations  between the mass of the SMBH and the mass of the host galaxy \citep{M98} and also between the mass of the SMBH and the stellar velocity dispersion of the host galaxy suggests that they  evolved together \citep{FM00,G09}. It is believed that SMBHs exist in the majority of nearby massive galaxies \citep{KH13}. They are the engines of active galactic nuclei (AGN) at high redshift.

The understanding of AGN is fundamental to understanding the co-evolution of SMBHs and their host galaxies. AGN are observed between the radio and gamma-ray wavebands. The observations in X-rays are useful to estimate the range of SMBH masses and Eddington ratios \citep{CC19}. An important fraction of AGN is obscured by dust where just the X-rays can escape from the system. On the other hand, some optical and near-infrared surveys have detected SMBHs even at $z>5$, such as ULAS J112'+0641 at redshift $z=7.085$ \citep{M11}, including the discovery of three quasars at $z>6$ by \citet{F03}. A very large number of SMBHs at high redshift are currently known; \citet{BA16} studied the properties of more than 100 quasars at $5.6 \lesssim z \lesssim 6.7$. The most distant AGN has been discovered at $z=7.5$ when the Universe was only $600\,$Myr old \citep{BA18}.

There are several theories about the formation of SMBH seeds and their evolution \citep{W19}.  One of the leading mechanisms in the literature is the direct collapse of massive gas cores in cold atomic cooling halos at high redshift \citep{BL03,W09,LS13b,LS15}. Simulations show that this process is only efficient if it is free of contaminants including metals, dust, and molecular hydrogen \citep{BL03, LO16}. Therefore, this process is very difficult to realize in nature, since even small amounts of dust can cause fragmentation \citep{O08,B16}. 

Another formation scenario involves the collisions and mergers of stars in a massive cluster with a high stellar density  \citep{D10,K15,S17,S19,R18,R20}. In particular in the context of primordial protostellar clusters, results from stellar evolution calculations suggest that collisions could play a large role, as the radii of these stars could be considerably enhanced depending on the accretion rate. For instance, for $10~$M$_{\odot}$ protostars the radii may range from $\sim30~$R$_\odot$, in the case of an accretion rate of $10^{-3}~$M$\rm_\odot yr^{-1}$,   up to $\sim500~$R$_\odot$,  in the case of   $1~$M$\rm_\odot~yr^{-1}$ \citep{H12,H13,H18,S13,W17}. Their results also show a strong and non-trivial dependence on the stellar mass. Some authors also propose a formation scenario via collisions and accretion \citep{B18,C20,T20,D20}. In this scenario the most massive star falls to the center of the cluster due to energy equipartition and momentum conservation during direct collisions.  This increases the rate of collisions and mergers since this new object has a larger collisional cross section than other stars in the cluster. Once several collisions have occurred, a very massive star (VMS) forms.  The VMS can have different final fates depending on its mass at the time of death. If the final mass is $M_f>260~$M$_{\odot}$ it will collapse into an intermediate-mass black hole (IMBH); for $150~$M$_{\odot}$ $<M_f<260~$M$_{\odot}$ it produces a pair-instability supernova \citep{H02}.

Fragmentation at low metallicity ($Z \geq 10^{-5}~$Z$_{\odot}$)  was often found to occur in disk-like structures \citep{C11b,D13}, and the presence of rotation in the spherical models of \citet{K66} also leads to a deformation in the outer zone of the cluster, including the appearance of a non-spherical distribution, such as a disk \citep{V12,LG87}. In this paper our  aim is to explore the evolution of flattened rotating star clusters that might form under such conditions. The effect of rotation is considered in the context of stellar populations and black hole binary mergers in Local Group galactic nuclei \citep{L16, L18}, though it is typically neglected when considering the formation of very massive objects via collisional processes. As a possible toy model we implemented the analytical solution of the  density and gravitational potential profile of \citet{MN75}, assuming the k-decomposition of \citet{S80} for the velocity in the azimuthal plane, to parameterize the amount of ordered versus unordered motion.  The evolution of these initial conditions was subsequently explored with {\sc Nbody6} \citep{A00}. The stellar radius is treated here as a free parameter to allow the potential application of our results to different scenarios and physical conditions.

In this paper we explore the evolution of a dense stellar Miyamoto-Nagai cluster including flattening and rotation. We describe the method to generate the initial conditions in section 2 and the methodology of our $N$-body simulations in section 3. We present our results in section 4. Finally, we discuss our results and conclude in section 5.
 
\section{Analytical model}\label{Analytical}

To compute the initial equilibrium state on the stellar systems that we study in this paper, we use the work of \citet{MN75}.  These authors derived potential and density profiles for stellar systems in cylindrical coordinates ($r,\varphi,z$), where $r$ is the radial coordinate, $\varphi$ is the azimuthal angle and $z$ is the vertical coordinate.  The profiles describe a flattened distribution of stars, which yields exact solutions of the Jeans equation for axisymmetric systems. The potential and density are given as the description of \citet{B87}:
\begin{equation}
\Phi(r,z)=-\frac{GM}{\sqrt{r^2+(a+\xi)^2}},
\end{equation}
\begin{equation}
\rho(r,z)=\left(\frac{Mb^2}{4\pi}\right)\frac{ar^2+(a+3\xi)(a+\xi)^2}{\xi^{3}(r^2+(a+\xi)^2)^{5/2}}.
\end{equation}
Here $\xi=\sqrt{z^2+b^2}$ and the shape of the disk is described by two parameters a and b which are constants with a dimension of length. If $a = 0$, it represents a spherical distribution of stars modeled by \citet{P11}.  If $b = 0$, it corresponds to a disk distribution modeled by \citet{K56}.  The chosen values for $a$ and $b$ determine the dimensionless quotient $c = b/a$. The higher the value of $c$, the rounder the system will be. A flattened system contributes more to the mean rotational velocity than the velocity dispersion of random motions \citep{NM76}; in other words, the second-order velocity moment is at least comparable in magnitude to the velocity dispersion.

Miyamoto and Nagai define the velocity dispersion as $(\sigma_r,\sigma_{\varphi},\sigma_z)$ (see also \citealt{P96}), which correspond to the cylindrical coordinates ($r,\varphi,z$). Following \citet{Binney1990}, they assume a distribution function that depends only on the isolating integrals corresponding to total energy $E$ and angular momentum $L_z$. This implies that averages like $\overline{v_R v_z}$, $\overline{v_R v_\Phi}$ and $\overline{v_\Phi v_z}$ are zero, implying that the velocity dispersion tensor is aligned with the coordinate system, and that the radial and axial velocity dispersions are equal (i.e., $\sigma_R=\sigma_z=\sigma$). The Jeans equation then becomes
\begin{eqnarray}
\frac{\partial}{\partial z}\left(\rho\sigma^2\right)&=&-\rho\frac{\partial}{\partial z}\Phi,\\
\frac{\partial}{\partial r}\left(\rho\sigma^2\right)+\frac{\rho\left(\sigma^2-\overline{v_{\varphi}^2}\right)}{r}&=&-\rho\frac{\partial}{\partial r}\Phi.
\end{eqnarray}
For the decomposition of the azimuthal velocity component $v_\varphi$, we adopt the k-decomposition by \citet{S80}
\begin{equation}
    \overline{v_\varphi}^2= k^2\left(\overline{v_\varphi^2}-\sigma^2\right),
\end{equation}
with $0\leq k \leq 1$ and then
\begin{equation}
\sigma_\varphi^2\equiv \overline{v_\varphi^2}-\overline{v_\varphi}^2=k^2\sigma^2+(1-k^2)\overline{v_\varphi^2}. 
\end{equation}
The case of $k=0$ corresponds to  no mean rotation, so all the flattening is due to the azimuthal velocity dispersion; $k=1$ corresponds to the so-called isotropic rotator solution; and  realistic cases with some net rotation have $0<k\leq1$.

Solving equation (3) with equations (1) and (2), we can write 
\begin{equation}
\rho\sigma^2 = \frac{Gb^2M^2}{8\pi}\frac{\left(a+\xi\right)^2}{\xi^2\left(r^2+\left(a+\xi\right)^2\right)^3}
,\end{equation}
where we recall that $\xi=\sqrt{z^2+b^2}$. Now, combining equation (7) with equation (4), we obtain
\begin{equation}
\rho\left(\overline{v_\varphi^2}-\sigma^2\right) = \frac{GM^2ab^2}{4\pi}\frac{r^2}{\xi^3\left(r^2+\left(a+\xi\right)^2\right)^3}.
\end{equation}
Combining equations (7) and (8), we obtain
\begin{equation}
\rho \overline{v_\varphi^2} = \frac{GM^2b^2}{4\pi\xi^2\left(r^2+\left(a+\xi\right)^2\right)^3}\left(\frac{ar^2}{\xi}+\frac{\left(a+\xi\right)^2}{2}\right).
\end{equation}

Our astrophysical model is described by the density and potential of Miyamoto-Nagai, assuming the k-decomposition of Satoh in the azimuthal plane.

\section{Simulations} \label{simulations}


We use {\sc Nbody6} \citep{A00} to run our simulations. The code includes an algorithm to regulate close encounters \citep{KS65}. The code also uses a Hermite fourth-order integrator with a spatial hierarchy to speed up computational calculations \citep{AC73}.

As our initial conditions describe a disk distribution, we calculate the initial stellar coordinates along the $z$-axis, then the radial coordinates, and finally the azimuthal coordinates. We define $L=a+b$, a factor to regularize the shape of the disk.

We use the inverse transform sampling method, for which we employ the enclosed mass at a given radius for a fixed $z$-coordinate.  We do this by integrating the density over rings on the $r\phi$-plane,
\begin{equation}
	I(r,z) = \int_0^{R} 2\pi r\rho(r)dr\\
	= \left(\frac{Mb^2}{2}\right)\left(\frac{aR^2+(a+\xi)^3}{\xi^3(R^2+(a+\xi)^2)^{3/2}}\right),
\end{equation}
and then we solve numerically for the values of the enclosed mass at a specific radius $R$ for a fixed array of $z$-coordinates, which is defined between $z_{min} = \log(10^{-8}\times L)$ and $z_{max}= \log(10^{2}\times L)$. Now using the trapezoid rule we integrate the enclosed mass over $z$ to obtain the total mass of the cylinder. Having generated a numeric lookup table, we now draw a new set of random numbers for $z$-coordinates and we interpolate these values with the enclosed mass that we obtain from our fixed array of $z$-coordinates. 

Having calculated the particle positions, we follow the description of the velocities given in section \ref{Analytical}, assuming a Gaussian distribution for the random variables. We now have the positions and velocities in the radial and $z$-coordinates, and we define the $\varphi$-coordinate as a set of random numbers between $0$ and $2\pi$. Finally, we use a rotation matrix to obtain the x and y coordinates, since {\sc Nbody6} uses Cartesian coordinates. We define the mass of the stars as the total mass of the cluster divided by the total number of stars.

We employ the initial conditions   outlined in section 2.  Initially, we set up a virialized system with a total mass of $M_{disk} = 10^5$ $\rm~M_{\odot}$.

We varied the initial number of stars as $N=10^3, 5\times10^3, 10^4$, and the mass of each star as $M_{star} = M_{disk}/N$. We varied the rotation as $k = 0,0.4,0.8,1$, and the initial stellar radius as $R = 50, 100, 500, 1000$ $\rm~R_{\odot}$. The radii are effectively considered here as a free parameter to be able to apply and relate the results obtained here to different scenarios. In particular the cases of larger radii are motivated by models for primordial protostars that have shown potentially very large radii for cases of $10-100$ $\rm~M_\odot$ stars \citep{H12,H13,H18,S13,W17}.

We consider collisions to occur when the stellar radii of two stars overlap ($d \leq R1 + R2$); when two stars satisfy this criterion, we replace them with a single new star. We also assume mass conservation during the collision. In reality, a small amount of mass should be lost \citep{D93,F05,G10}, but it does not represent a significant change in the final mass of the collision product. The parameters of the new star after the merger are as follows:
\begin{equation}
	M_{new}=M_1+M_2,
\end{equation}
\begin{equation}
	R_{new}=R_1(1+M_2/M_1)^{\frac{1}{3}}.
\end{equation}
   We assume that the new object quickly reaches a new internal equilibrium, radiating away any excess energy deposited in the post-collision product and returning to hydrostatic and thermal equilibrium.

\begin{figure}[!ht]
    \centering
    \includegraphics[width=\hsize]{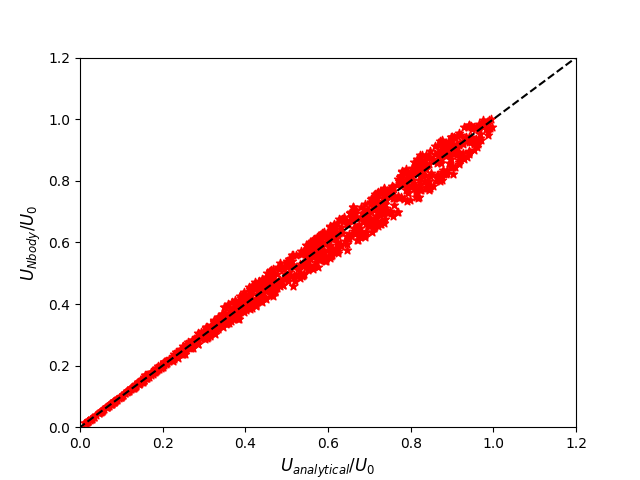} 
    \caption{Plot of the N-body potential energy $U_{Nbody}$ of a particle. $U_0$ is the normalization of the potential energy in the center of the cluster and $U_{analytical}$ is the analytical value of the potential energy at that position. This example corresponds to the flatter cluster A, with rotation $k=1$, and number of stars $N=10^3$.}
\end{figure}

In Fig. 1, we summarize the reliability of our initial conditions showing that the N-body potential energy obtained from the particle discretization is close to the analytical value of the potential energy.

\section{Results}


In this section we summarize our main results, showing the time evolution of the  Lagrangian radii both in the radial and vertical direction at $90\%$, $50\%$, and $10\%$ of the enclosed mass. We analyze in particular how rotation and flattening affect the rate of collisions and stellar ejections from the system.


\subsection{Time evolution and energy equilibrium}


\begin{table}
\caption{Values for $a$, $b$, and $c$.}
\centering
\begin{tabular}{cccc}
\hline\hline
cluster name& $a$ $\rm~[pc]$ & $b$ $\rm~[pc]$ & $c$\\
\hline
A & 0.06 & 0.04 & 0.7 \\
B & 0.04 & 0.06 & 1.5 \\ 
C & 0.01 & 0.09 & 9.0\\ 
\hline
\end{tabular}
\tablefoot{$a$, $b$, and $c$ are parameters that define the shape of the model cluster. A higher value of $c$ corresponds to a more spherical system.}
\end{table}

\begin{figure}[!ht]
\centering
\includegraphics{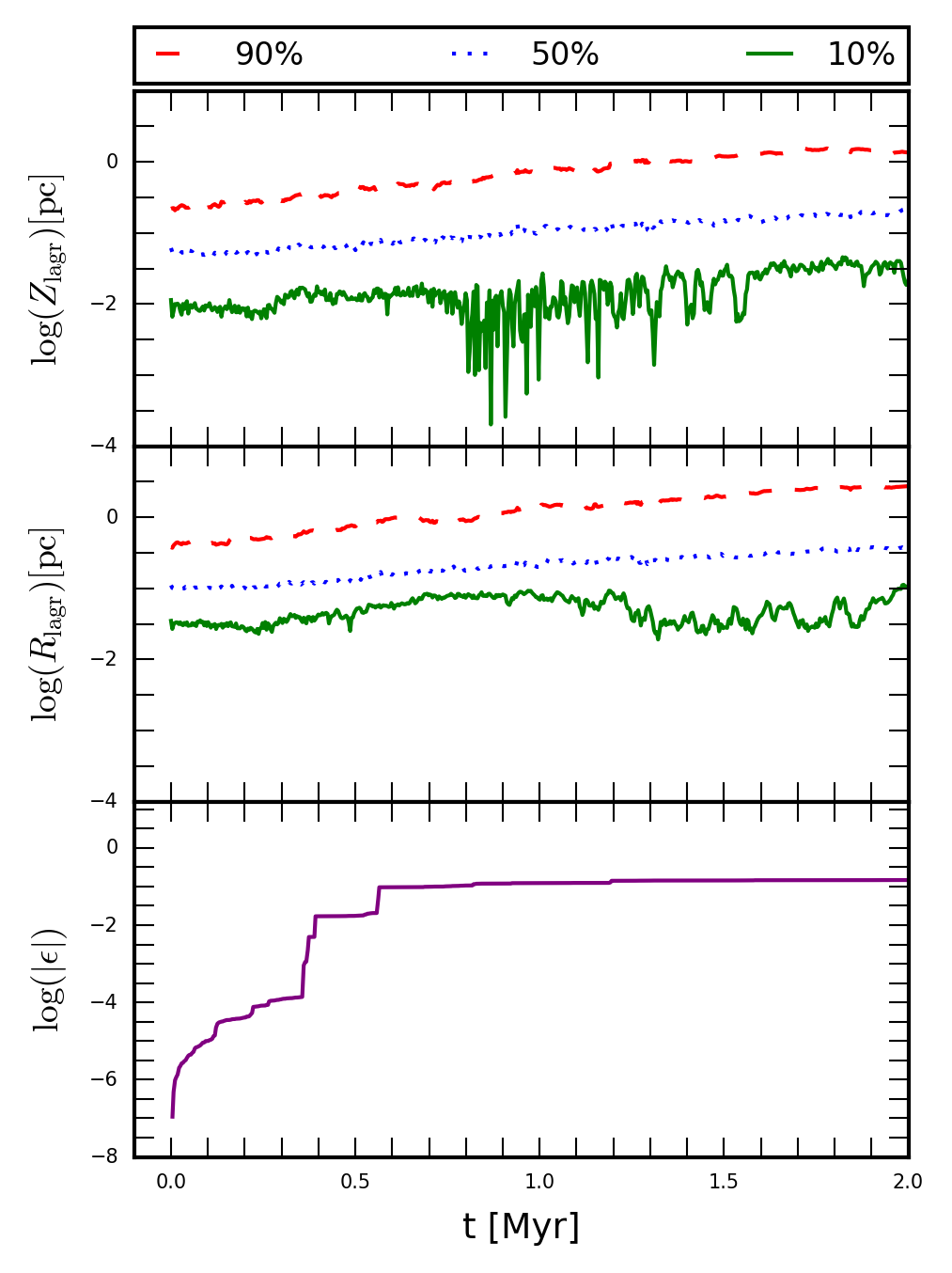}
\caption{Evolution over time for cluster C without rotation, with $N=10^3$ stars, where each star has a radius of $R = 100$ $\rm~R_{\odot}$. Top panel: Lagrangian coordinate in the vertical direction for $90\%$, $50\%$, and $10\%$ of the enclosed mass as a function of time. Middle panel: Lagrangian radii for the $90\%$, $50\%$, and $10\%$ mass intervals as a function of time. Bottom panel: 
 Relative cumulative energy error $(\epsilon)$ (equation~(\ref{energyerror})) is given as a check of the energy conservation as a function of time.}
\label{Lagrange}
\end{figure}

We explore three types of clusters, from a  flatter to a more spherical distribution, as outlined in Table 1. Due to the flattened geometry of the system, we express the Lagrangian radii with respect to both variables $R$ and $z$. The Lagrangian radii in the radial direction at $90\%$, $50\%$, and $10\%$ of the enclosed mass is shown as a function of time in Fig.~\ref{Lagrange} for cluster C without rotation ($k=0$) with $N=10^3$ and $R=100$ $\rm~R_{\odot}$. In the radial plane we see a slow expansion of the Lagrangian radii at $50\%$ and $90\%$ of the enclosed mass; the inner zone shows rapid fluctuations at all times, also with a slow expansion that decays after $1.2\,$Myr. 

For comparison, we also show the Lagrangian coordinate in the vertical direction $z$. These coordinates are obtained considering all particles independent of their radial coordinate, and determining the amount of mass between the $x-y$ plane and a fixed coordinate $z$. When this mass corresponds to a fraction $\alpha$ of the total mass, we refer to it as the corresponding Lagrangian coordinate in z-direction for fraction $\alpha$. The Lagrangian $z$-coordinates of the outer and middle zone of the cluster (i.e., at $50\%$ and $90\%$ of the enclosed mass) expand by about a factor of $5$ over two million years. However, the inner Lagrangian radius (at $10\%$ of the enclosed mass) slightly shrinks over the first $0.15\,$Myr, followed by a moderate expansion of about a factor of $2$ over the next $0.5\,$Myr. At $0.7\,$Myr, the fluctuations become deeper and show an oscillatory behavior, while after $1.5\,$Myr the inner zone of the system becomes more stable, similar to the beginning. 

The relationship between the Lagrangian radii in the $z$ and $R$ directions determines the flatness of the distribution. In the case of a uniform geometry, the ratio of these quantities would ideally be constant, which we find to be approximately the case for the $50\%$ and $90\%$ Lagrangian radii, while the innermost radius appears to  change with time. In a more general sense, instead of Lagrangian radii, Lagrangian volumes that correspond to certain fractions of enclosed mass can also be considered.

As a check of the energy conservation in the system, we also show the relative cumulative energy error 
\begin{equation}
	\epsilon = \sum_i\left |\Delta E_i/E\right|,\label{energyerror}    
\end{equation}
 with $E$ being the initial total energy and $\Delta E_i$ the change in total energy in time steps of  $i$. We find that the increase in $\epsilon$ is essentially driven by the growth of the most massive object via mergers. While our treatment of mergers ensures the conservation of linear momentum, the sum of kinetic and gravitational energy is not expected to be conserved during the collision, and the change in $\epsilon$ appears closely related to the growth of the most massive object.

\begin{figure}[!ht]
\centering
\includegraphics{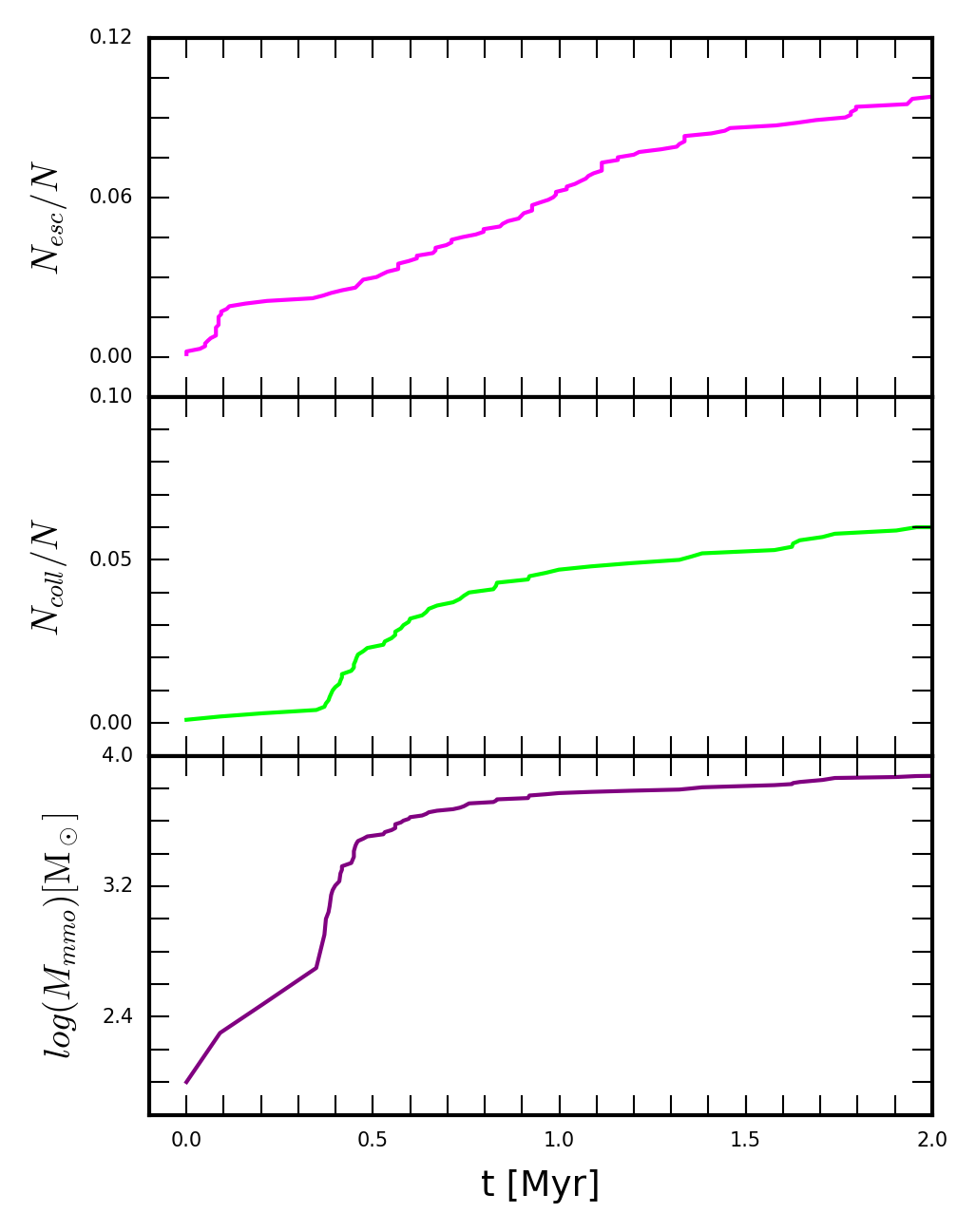}
\caption{Same cluster  as in Fig.~\ref{Lagrange}. Top panel: Normalized number of escapers. Middle panel: Normalized number of collisions with the most massive object. Bottom panel: Mass of the most massive object.}
\label{esc-coll-mmo}
\end{figure}

In Fig.~\ref{esc-coll-mmo} we show the time evolution of cluster C without rotation (i.e., $k = 0$) with $N= 10^3$ and $R = 100$ $\rm~R_\odot$. In the top panel we expose the cumulative number of escapers normalized by the total number of stars. In the first $0.1\,$Myr there are approximately 10  stars ejected from the system, followed by a smooth expansion to 100   ejected stars at $2\,$Myr. In the middle panel we show the accumulated number of collisions with the most massive object normalized by the total number of stars. Up to $0.4\,$Myr there are very few collisions, then in the next $0.5\,$Myr we can see an increase in the stars of around $5\%$,  which means around 50 collisions, followed by an increase of $1\%$ in the fraction of collisions until $2\,$Myr. Finally, the bottom panel displays the logarithm of the mass of the most massive object.  Until $0.4\,$Myr the growth of the most massive object is slow, the following $0.5\,$Myr shows the fastest growth, after $0.9\,$Myr the growth is very slow. There are clear correlations between the shape of the curves in the middle and bottom panels.


\subsection{Collisions, escapers, and the most massive object}


In the following subsection we discuss the number of stars that escape from the system, the number of collisions with the most massive object both normalized by the total number of stars and also the logarithm of the mass of the most massive object.

The fraction of escapers appears almost constant for  factors of rotation $k=0, 0.4, 0.8$. For the higher value of $k=1$, there is a larger fraction of stars which remains inside the system, suggesting that rotation may promote the retention of stars in the cluster and inhibit stellar escapes. We note that retaining the stars does not give rise to more collisions, as the overall motion is more ordered in systems with a higher degree of rotation, so the conservation of angular momentum prevents them from falling towards the center. The growth of the most massive object is directly related to the number of collisions with the most massive object. The final mass of the most massive object is slightly reduced in systems with maximum rotation $k=1$ compared to systems with less rotation $k= 0.4, 0.8$ or without rotation $k=0$. 

On the other hand, the escaper fraction and the number of collisions increases with larger initial stellar radii. This happens because larger stellar radii mean larger cross sections, so the possibility of an encounter between stars is higher, which means more gravitational interactions, which also leads to a higher number of escapers. As the growth of the most massive object is directly linked to the number of collisions, its final mass also increases for larger initial stellar radii.

\begin{figure*}[htp!]
  \centering
    \includegraphics{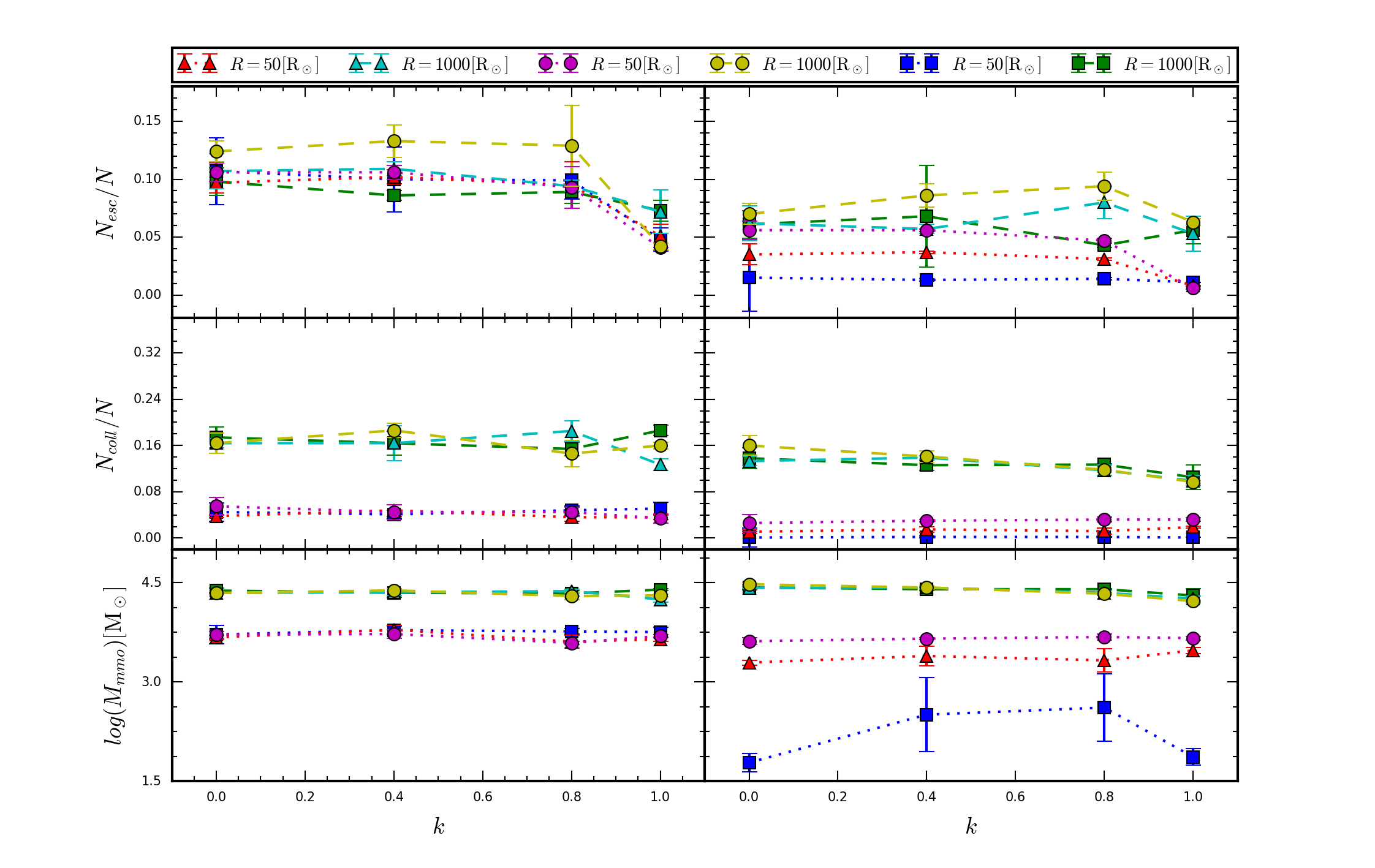}
    \caption{Evolution of type A, B, and C clusters represented by a triangle, a circle, and a square, respectively, with $ N = 10^3 $ stars on the left and $ N = 10^4 $ on the right. We show the initial stellar radii $R = 50, 1000$ $\rm~R_{\odot}$ as a function of the $k$-factor. Top panel: Normalized number of escapes. Middle panel: Normalized number of collisions with the most massive object. Bottom panel: Mass of the most massive object.}
    \label{Ns_vs_k}
\end{figure*}

In Fig.~\ref{Ns_vs_k} we display the total number of escapers normalized by the total number of stars, the total number of collisions with the most massive object normalized by the total number of stars, and the mass of the most massive object, for total numbers of stars $N = 10^3$ and $N = 10^4$. We provide the results of simulations with two different initial stellar radii, $R=50$, $1000\rm~R_{\odot}$, for the type A, B, and C clusters as a function of the $k$-factor.

In the top left panel the escape fraction remains more or less constant for  $k=0, 0.4, 0.8$, while it decreases for $k=1$, suggesting that rotation may promote the retention of stars in the cluster and reduces the number of stellar escapes. In the case of the large initial stellar radius $R  = 1000$ $\rm~R_{\odot}$ the differences between the clusters become more visible, and the type C clusters, which is the rounder shows fewer escapers. The type A clusters, which is the flattest experiences more escapers than type B clusters. In the case of small initial stellar radii $R = 50$ $\rm~R_{\odot}$ the behavior of the different configurations appears quite similar. In the top right panel for the larger initial stellar radius $R  = 1000$ $\rm~R_{\odot}$, we find that the escape fraction in type A and B clusters increases with increasing factor of rotation until $k=0.8$, then shows a decrease at $k = 1$. For type C clusters there are more escapers for the lower factors of rotation $k = 0, 0.4$ than for the higher values of rotation $k=0.8,1$. On the other hand, for the smallest initial stellar radius $R  = 50$ $\rm~R_{\odot}$, the fraction of escapers is higher in type B clusters, followed by type A and C clusters, at least for the three first factors of rotation $k=0,0.4,0.8$. Nevertheless, the fraction of escapers is quite similar for $k<1$. For the last factor of rotation $k=1$, all the configurations show the lowest fractions of escapers. 

The middle left panel shows the total number of collisions with the most massive object normalized by the total number of stars. For the larger initial stellar radius $R  = 1000$ $\rm~R_{\odot}$ the number of collisions with the most massive object is quite similar for each configuration; the differences between the configurations increase with increasing $k$-factor. While the initial stellar radius changes by orders of magnitude, the fraction of collisions changes only by roughly a factor of 2. In the middle right panel  the number of collisions with the most massive object is quite insensitive to geometry and $k$-factor. The differences between type B and C clusters are larger since the fraction of collisions in type C clusters are close to zero, while in type B clusters it reaches values near 0.08, which means around 800 collisions more. The amount of rotation does not appear to affect the fraction of collisions very significantly.   

The bottom left panel shows the mass of the most massive object as a function of the $k$-factor, which is quite insensitive both to the shape and the $k$-factor, for small and large initial radii $R  = 50, 1000$ $\rm~R_{\odot}$. This suggests that rotation does not affect collisions at least in the case of $ N =  10^3 $ stars. In the bottom right panel, for the case of the largest stellar radius, the mass of the most massive object is again rather independent of geometry and $k$-factor. There is a small decrease in  the final mass for  $k = 1$. On the other hand, for the smaller initial stellar radius $R  = 50$ $\rm~R_{\odot}$,  the final mass varies more strongly. The rounder type C clusters shows the least massive object, suggesting that the flattening increases the number of collisions due to the higher density it produces. 

\begin{figure*}[htp!]
  \centering
    \includegraphics{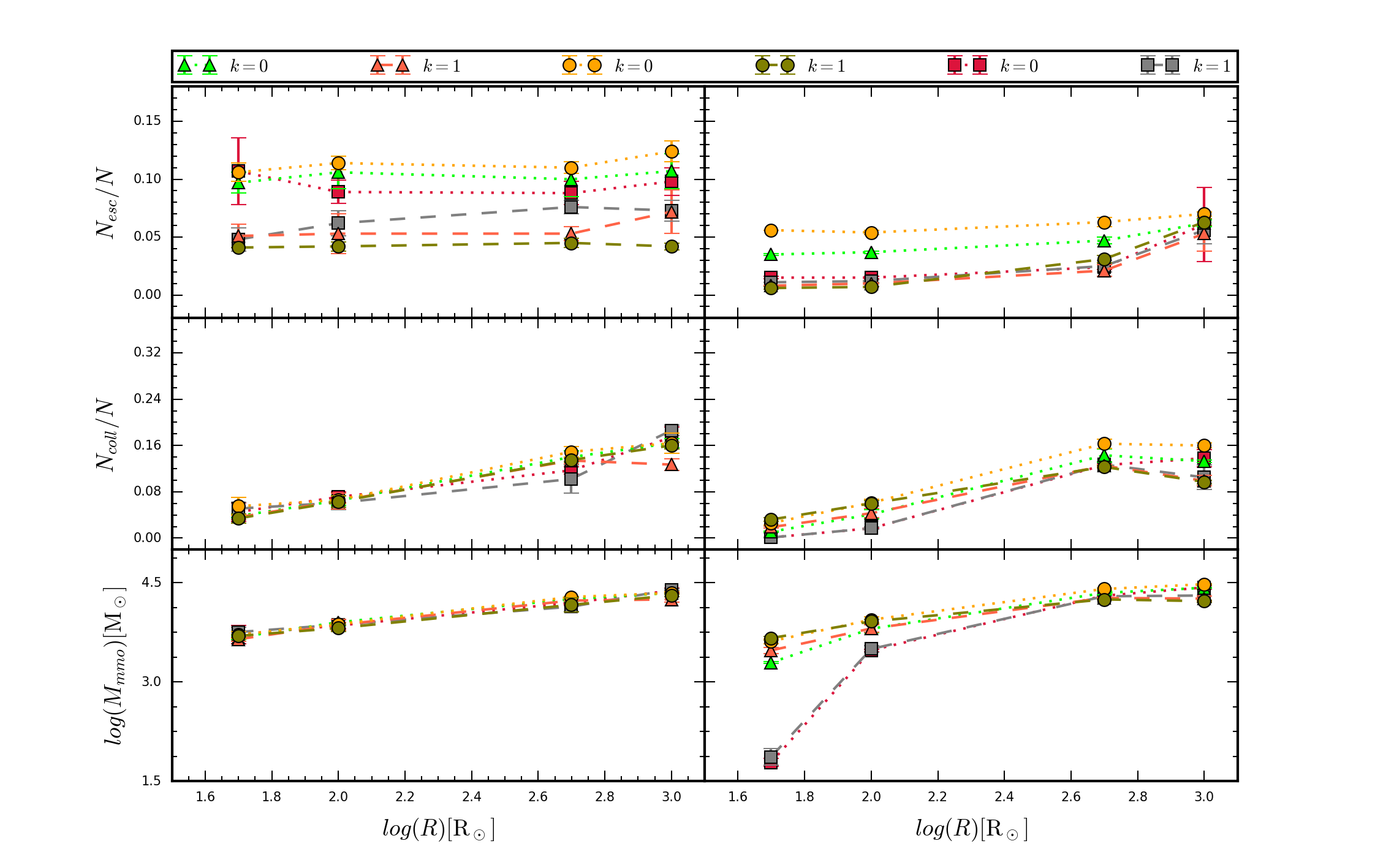}
    \caption{Evolution of type A, B, and C clusters. Symbols, and panels are the same as in Fig.~\ref{Ns_vs_k}, with $ N = 10^3 $ stars on the left and $ N = 10^4 $ on the right. We show the results for factors of rotation $k=0, 1$ as a function of the initial stellar radii $R$.}
    \label{Ns_vs_R}
\end{figure*}

In Fig.~\ref{Ns_vs_R} we show the same quantities given in Fig.~\ref{Ns_vs_k} as a function of the initial stellar radius, for two factors of rotation $k=0, 1$. We show in the top left panel the fraction of escapers which is clearly larger for the clusters without rotation. For $ k = 0 $, the simulations show small differences in the  fraction of escapers in simulations with the smaller initial stellar radius $R = 50$ $\rm~R_{\odot} $ , while for the larger initial stellar radius $R  = 1000$ $\rm~R_{\odot}$ the fraction of escapers shows an increase  in type A and B clusters, and type C clusters shows fewer escapers. All clusters show the largest fraction of escapers for an initial stellar radius of $R = 1000$ $\rm~R_{\odot} $. On the other hand, the systems with rotation ($k=1$) show the smallest fraction of escapers. Considering the geometry, type B clusters has the smallest fraction of escapers, followed by type A clusters, and finally the rounder type C clusters  shows the highest number of escapers. Type B clusters shows a similar number of escapers for the different stellar radii $R=50, 100, 500, 1000$ $\rm~R_\odot$; type A clusters shows a similar number of escapers for the first three stellar radii $R = 50, 100, 500$ $\rm~R_{\odot} $ and  an increase for the largest stellar radius $R=1000$ $\rm~R_\odot$. type C clusters shows a smooth increase in the escape fraction up to $R = 500$ $\rm~R_{\odot} $, followed by a slight decrease for the largest stellar radius $R=1000$ $\rm~R_\odot$. The top right panel shows that the clusters with rotation show a lower escape fraction than the clusters without rotation. We note that for type C clusters the number of escapers is quite similar  for the configurations with and without rotation, while the type A and B clusters show more differences between the escape fractions.

The middle left panel  shows a clear correspondence between the increase in the number of collisions and the increase in the initial stellar radii, except for the configuration of type A clusters with rotation, which shows a small decrease for the case of the largest initial stellar radius $R  = 1000$ $\rm~R_{\odot}$. In the middle right panel all configurations show an increase in the number of collisions against the increments of the initial stellar radii up to $R = 500$ $\rm~R_{\odot} $, when the system with rotation shows a decrease in the number of collisions, while the systems without rotation maintain a constant number of collisions between the largest initial stellar radii $R = 500, 1000$ $\rm~R_{\odot} $.

The bottom left panel shows the mass of the most massive object. The final mass of the most massive object appears again to be rather insensitive to the details of the  configurations examined here. In the bottom right panel there are no big systematic differences between the final mass of the most massive objects in a system with or without rotation. However, there is a difference in the final mass of the most massive object related to the shape of the cluster, since the rounder type C clusters shows the least massive object at least for the smallest initial stellar radii $R =50, 100$ $\rm~R_{\odot}$. On the other hand, for the largest initial radii $R =500, 1000$ $\rm~R_{\odot}$ we find a similar final mass of the most massive object for each cluster. In addition, type B clusters shows the most massive object of all configurations.

In summary, we find that rotation described via the $k$-factor  does not lead to great differences in the total number of collisions with the most massive object. We also found that rotating systems show fewer escapers of stars. However this does not lead to more collisions since the rotating systems are more ordered, meaning that  their stars have stable orbits, which avoid   sinking to the center. These results are based on the average of three simulations for the same unique configuration to obtain reliable statistics.


\subsection{Total number of collisions}


In the previous subsection we focused on the number of collisions with the most massive object. However,  the total number of collisions between any two stars in the system is also relevant as it changes the stellar mass distribution and may lead to the formation of additional massive objects (which may or may not later merge with the most massive one). 

\begin{figure*}[htp!]
  \centering
    \includegraphics{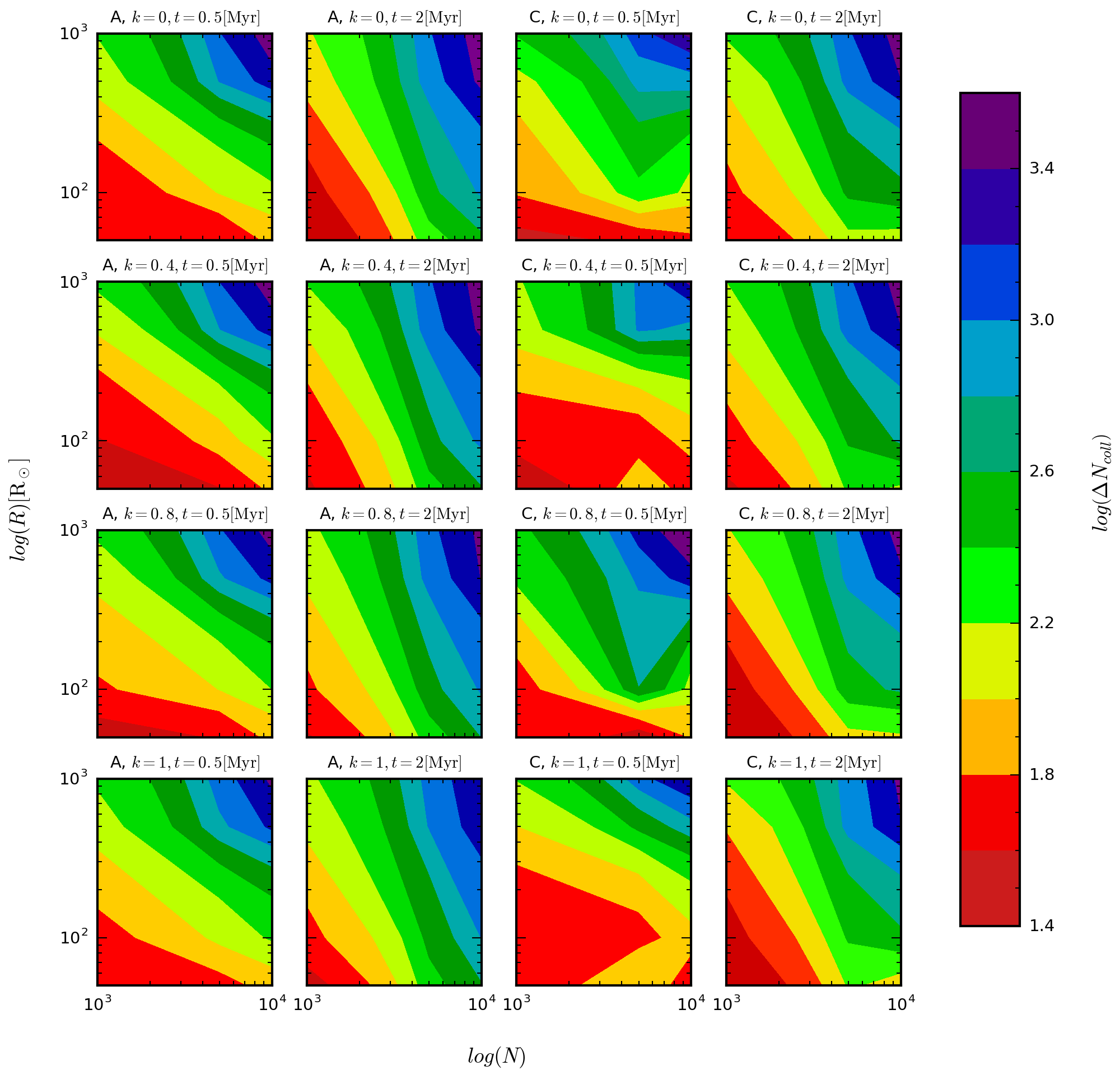}
    \caption{Total number of collisions $N_{\rm coll}$ in type A and C clusters as a function of the number of stars and the initial stellar radii for $k$-factors  $k=0,0.4,0.8,1$, until $0.5\,$Myr (columns 1 and 3) and $2\,$Myr (columns 2 and 4).}
    \label{4x4}
\end{figure*}

To show how the total number of collisions depends on our main parameters, we built color diagrams representing the total number of collisions (see Fig.~\ref{4x4}). We provide the total number of collisions for different total numbers of stars and different initial stellar radii as we expect the number of collisions to be regulated by both quantities. There are 16 panels,   each one representing a different factor of rotation, geometry, and evolution time. The first row of panels is for $k=0$, the second row is for $k=0.4$, the third row is for $k=0.8$ and the last row is for $k=1$. We recall that for $k=0$ all stellar motions are random, while $k=1$ corresponds to an ordered motion dominated by rotation. 

Type A clusters are the flattest, and therefore the densest; type C clusters are the roundest. We find that the flatter clusters show more collisions than the rounder ones;  for instance, the central number density of type A clusters are higher, which means that there are more star encounters. In all panels of each figure, we can see from left to right how the number of collisions increases as a function of the total number of stars from $N =10^3$ to $N =10^4$. We show from bottom to top the increase in the number of collisions as a function of the initial stellar radius from the smallest stellar radius $R=50$ $\rm~R_\odot$ to the largest $R=1000$ $\rm~R_\odot$. The number of collisions is regulated by the number of stars and their initial stellar radii. This means that   in a system with a greater number of stars there are more collisions, and also means that stars with larger initial stellar radii have larger cross sections, which implies an increase in the total number of collisions. 

Going from the first to the second columns, we see the evolution when going from $0.5$ to $2$ million years  in the total number of collisions in type A clusters. This increase is represented by the extension of the blue-purple zone toward the green zone. We note that the blue-purple zone represents the largest initial stellar radii and the highest numbers of stars, and the total number of collisions is regulated by both quantities. On the other hand, the factor of rotation does not contribute to the total number of collisions. In general, all the panels show a  similar behavior for the number of collisions; for $ k = 1 $ we find the smallest fraction of collisions. This is clear considering the values in the upper right and lower left corners and in comparison with the other panels for different factors of rotation, which lead to a higher fraction of collisions. The total number of collisions is regulated by the total number of stars and their initial stellar radii.  The highest total number of collisions is obtained for the highest number of stars and the largest initial stellar radii. In the third and fourth columns, we can see the  total number of collisions of type C clusters both at $0.5$ and $2$ million years. At $0.5\,$Myr the factor of rotation appears to contribute a little to the total number of collisions; this is not visible at  $2\,$Myr, where the highest numbers of collisions appear in the upper right corner. The larger  number of stars and the larger initial stellar radii show a higher total number of collisions. The rotation factor $k$ slightly reduces the collisions due to ordered motions since the rotation factor $ k = 1 $ shows the smallest fraction of collisions. 

In brief, we find that changes in the rotation factor do not lead to great differences in the total number of collisions. On the other hand, the flatter type A clusters shows a greater number of collisions than the rounder type C clusters. This can be explained because type A clusters are denser than type C clusters. Furthermore, the total number of collisions increases with an increase in the number of stars and in their initial stellar radii. These results are based on the average of three simulations for the same unique configuration to obtain reliable statistics.


\subsection{Impact of flattening}


In this subsection we summarize our main results concerning the impact of flatness on the number of collisions. To explain this we display a plot of the total number of collisions normalized by the total number of stars against the central number density ($\eta$) in type A,B, and C clusters.

\begin{figure}[!ht]
  \centering
    \includegraphics{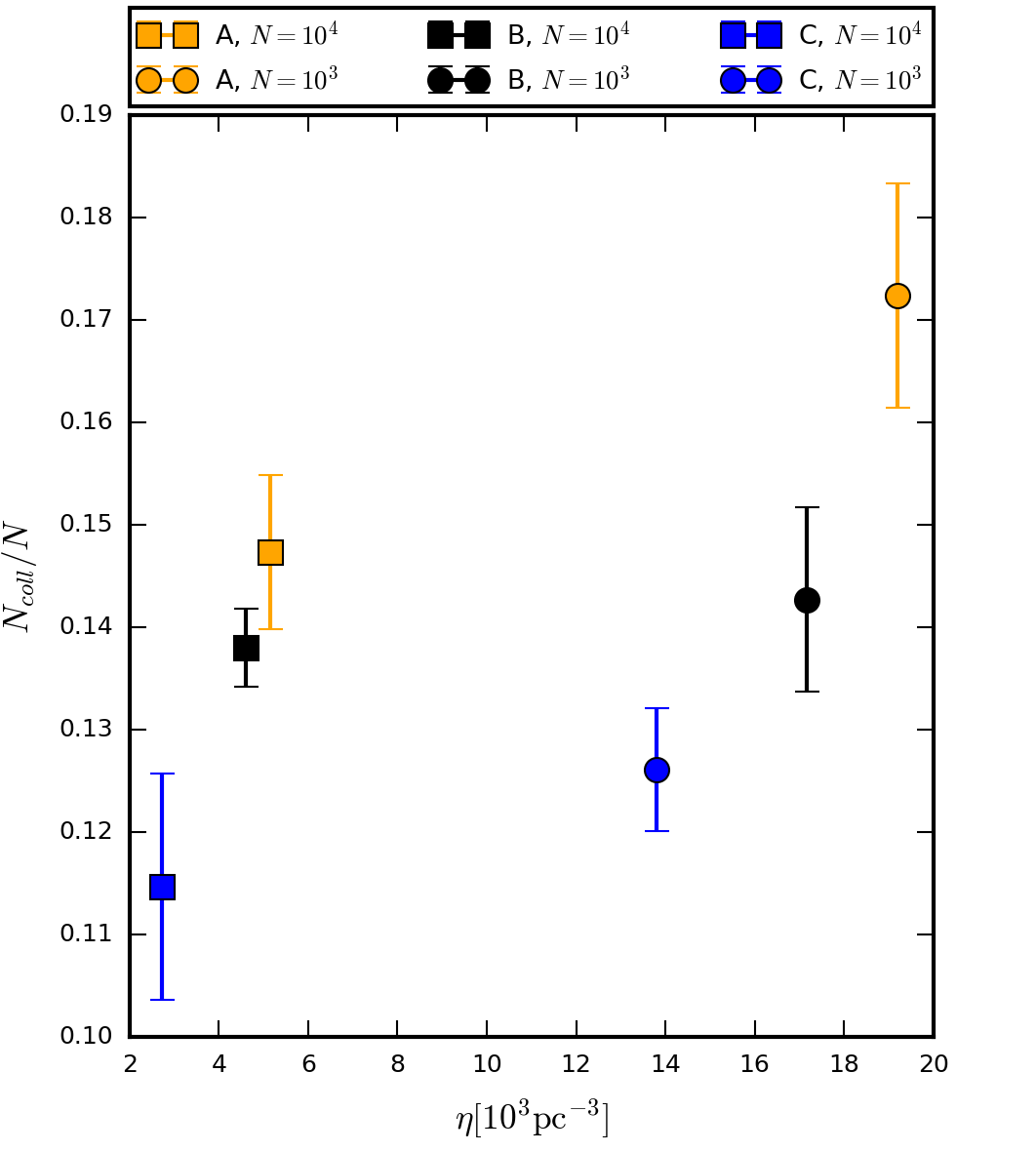}
    \caption{Number of collisions $N_{\rm coll}$ normalized by the total number of stars against the central numeric density of the  systems without rotation and with initial stellar radii of $R=500$ $\rm~R_\odot$. The squares are for systems with $N=10^3$; the circles are for systems with $N=10^4$. The orange symbols represent the flatter type A clusters, the black symbols type B clusters, and the blue symbols  the  rounder type C clusters.}
    \label{Ncoll_Ndensity}
\end{figure}

In Fig.~\ref{Ncoll_Ndensity} we show the total number of collisions normalized by the total number of stars $N=10^3,10^4$ as a function of the central number density in type A, B, and C clusters. Systems normalized by $N=10^3$ (squares) and $N=10^4$ (circles) are shown. Type A clusters are  the flattest and thus the densest, and shows the highest number of collisions; type C clusters are the roundest, and shows the smallest number of collisions. We note that the circles show a larger central number density than the squares since they have more stars, and they also show a higher number of collisions. There is a fairly clear correlation between the central number density and the total number of collisions. A higher central number density means a higher number of collisions, for both configurations of $N=10^3$ and $N=10^4$ stars. 

In summary,  flattening is one of the main parameters that regulates the total number of collisions together with the initial stellar radii and the total number of stars. The flattening contributes to the number of collisions since their stars are confined to a thinner region, which means that they are denser systems, so an encounter between the stars is more probable. Then we conclude that the number of collisions is regulated by the flattening. These results are based on the average of three simulations for the same unique configuration to obtain reliable statistics.

\section{Discussion and conclusions}


In this paper we investigated the evolution of flattened rotating star clusters to measure their influence on rate of stellar collisions. In order to parameterize both the ordered and random motion, we implemented the analytical solution of the density and the potential profile of \citet {MN75}, assuming the $ k$-decomposition of \citet {S80} for the velocity in the azimuthal plane.

We pursued a detailed parameter space study with our implementation of a Miyamoto-Nagai disk.  Our results quantify how rotation and flattening affect the evolution of the clusters, with a focus on the rate of stellar collisions. We analyzed the influence of different initial numbers of stars and initial stellar radii on the collision rate and the rate of formation of the most massive object. We measured the number of escapers, the number of collisions with the most massive object, the growth of the mass of the most massive object, and the total number of collisions (i.e., including collisions that do not necessarily involve the most massive object). All the systems with or without rotation show very similar total numbers of collisions when the total number of particles and particle masses and radii are the same. 

The systems with rotation retain more stars, and have fewer escapers.  This occurs because the ordered motions due to angular momentum conservation decrease the rate at which long-range interactions make stars become unbound from the cluster. The increased fraction of stars that are kept within the system does not lead to a higher rate of collisions.  We find that the systems with more rotation experience a lower collision rate with the most massive object, since they keep their angular momentum, preventing them from sinking to the center where the most massive object resides.

As expected, we find that the number of collisions increases with initial stellar radius, which we varied from $50-1000$ $\rm~R_\odot$.  This occurs because these stars have a larger collisional cross section, such that encounters between stars are more common. The number of collisions also increases with the initial number of stars (from $ N=10^3$ to $N =  10^4$) since the collision rate increases with $N$.  This result is consistent with the simulation results of \citet{R18}. Also for the Miyamoto-Nagai disk, we find these to be the main parameters that regulate the collisions, very similar to the case of a simple Plummer sphere. There are additional effects due to the geometry and the amount of rotation (which we varied as independent quantities), though in comparison their impact is smaller. Even when comparing rotating with non-rotating models, the difference is at most about  $20\%$. For the geometry, in comparison with the same number of stars and the same extent in the radial direction, then different shapes of the cluster  changed the collision rate by at most $25\%$ in the case of the flattest Miyamoto-Nagai distribution with scale height corresponding to about $10\%$ of the radius. 

After two million years, a typical lifetime of very massive stars, we find that under the most extreme conditions,  up to $4000$ collisions may occur, assuming $N=10^4$ stars and stellar radii of $1000$ $\rm~R_\odot$ for a cluster where initially about $90\%$ of the mass is confined in a radius of $0.3$ $\rm~pc$. Correspondingly, the $\rm~M_{MMO}$ may then reach up to $40000$ $\rm~M_\odot$, relatively independent of the specific geometry and rotation rate. For a more moderate case with initial stellar radii of $100$ $\rm~R_\odot$, the number of collisions is about $630$, and the mass after  two million years, thus about $6300$ $\rm~M_\odot$. The mass estimates after that time  should be regarded as an upper limit as we do not consider mass loss during collisions or due to stellar winds. However, we note that the runaway growth reported here would continue as long as the cluster remains dense and bound, so any conclusion on the final mass of the most massive object is premature and requires new sets of simulations that include stellar evolution and a realistic prescription of the gas dynamics of the cluster and its galactic environments on a larger scale.

A potentially relevant effect that we did not explore here concerns the consequences of mass segregation, which in particular for clusters more massive than the ones considered here may become increasingly more relevant. Mass segregation may either be due to dynamical effects in the cluster \citep{M07,A09,A10,Y11,P16,D17}, or it can be primordial, due to the preferential formation of more massive stars towards the center of the cluster \citep{Z82,M96,E01,K01,B01,B06}. The effects of mass segregation dynamically occurring in a cluster  are incorporated in our models, and the formation of a very massive object can be considered  an extreme case of dynamical mass segregation, though  they could be enhanced if the clusters are more massive. The presence of primordial mass segregation, on the other hand, could favor collisions in the center of the cluster, even in the presence of rotation, and thus further support the formation of massive objects in very massive systems. Potentially it is thus conceivable that the results which we have derived here will depend on the mass of the cluster due to the mass segregation process.


\begin{acknowledgements}


We are very grateful for funding grants by the Conicyt PIA ACT172033, Fondecyt regular (project code 1161247) as well as the BASAL Centro de Astrof\'isica y Tecnolog\'ias Afines (CATA) AFB-170002. MZCV, DRGS and NWCL acknowledge financial support from Millenium Nucleus NCN19$\_$058 (TITANs). These resources made the presented work possible, by supporting its development. This project was also supported by funds from the European Research Council (ERC) under the European Union’s Horizon 2020 research and innovation program under grant agreement No 638435 (GalNUC). NWCL gratefully acknowledges the generous support of a Fondecyt Iniciac\'ion grant 11180005. BR acknowgledges funding through ANID (CONICYT-PFCHA/Doctorado acuerdo bilateral DAAD/62180013) and DAAD (funding program number 57451854).
RSK acknowledges support from the Deutsche Forschungsgemeinschaft (DFG) via the Collaborative Research Center (SFB 881, Project-ID 138713538) ''The Milky Way System'' (sub-projects A1, B1, B2 and B8) and from the Heidelberg cluster of excellence (EXC 2181 - 390900948) ''STRUCTURES: A unifying approach to emergent phenomena in the physical world, mathematics, and complex data'', funded by the German Excellence Strategy. He also acknowledges funding from the European Research Council in the ERC Synergy Grant ''ECOGAL -- Understanding our Galactic ecosystem: From the disk of the Milky Way to the formation sites of stars and planets'' (project ID 855130). 

\end{acknowledgements}

%
%

\end{document}